\documentclass[12pt]{article}
\usepackage{epsfig}
\title{Symmetry algebras for superintegrable systems}
\author{C. Gonera\thanks{supported by KBN grant 5 P03B06021},  
P. Kosi\'nski$^*$, M. Majewski$^*$, P. Ma\'slanka$^*$\ 
\\Department of Theoretical Physics II \\
University of {\L}\'od\'z \\
Pomorska 149/153, 90 - 236 {\L}\'od\'z/Poland.}
\date{}

\begin{document}
\maketitle
\begin{abstract}
It is shown that the symmetry algebra of quantum superintegrable
 system can be always chosen to be $u(N),\;N$\ being the number of degrees of freedom.
\end{abstract}

\newpage

Classical dynamical system of $N$\ degrees of freedom is called maximally superintegrable if it admits $2N-1$\ independent,
globally defined integrals of motion. They form the complete set of integrals and any other integral of motion 
can be expressed in terms of them. In particular, the Poisson bracket of two basic integrals,
 being again a constant of motion,
is  (in general, nonlinear) their function. 
We conclude that the integrals of motion for maximally superintegrable system
form a finite $W$-algebra \cite{b1}, \cite{b2}.

An interesting question arises whether this algebra can be linearized
 to Lie algebra by a judicious choice of basic integrals. 
The answer to this question, for a large class of confining 
systems, admitting action-angle variables is positive \cite{b3} (see Ref. \cite{b4} for older
papers on the subject, both in classical and quantum theory).
 Moreover, assuming that \underline{all} action variables do appear
in the hamiltonian, the Lie algebra obtained is universally $u(N)$. 
The price one has to pay for this result is that the resulting
integrals are usually quite complicated functions of basic dynamical variables.

The aim of the present note is to extend this result to the quantum case.

The first problem which arises when passing to the quantum case is the very definition of quantum superintegrable system. 
We shall use the following one which seems to cover most intresting cases. 
Assume we have a set of commuting selfadjoint operators $\hat{I}_k,\;k=1,\dots ,\;N$\ (quantum actions) such that:\\
(i) the spectrum of each $\hat{I}_k$\ consists of the eigenvalues of the form $n_k+\sigma_k$, where $n_k\in N$\ 
is any natural number and $\sigma_k \in {\bf R}$\ is fixed.\\
(ii) the common eigenvectors of all $\hat{I}_k,\;\mid n>,\;n\equiv (n_1,\;\dots ,\;n_N)$, span the whole space of states\\
(iii) the hamiltonian can be written as
\begin{eqnarray}
\hat{H}=H(\sum_{k=1}^Nm_k\hat{I}_k)\label{w1}
\end{eqnarray} 
with $m_k\in N,\;k=1,\;\dots ,\;N$; we shall also assume that all $m_k\neq 0$. Let us also note that we can assume that
$m_1,\;\dots ,\;m_N$\ have no common divisor except one.

We believe the above definition is flexible enough to cover known cases. Sometimes it may appear necessary to take an orthogonal
sum of few Hilbert spaces described above to cover the whole space of states (the simplest example semms to be the planar Kepler problem).

The vectors $\mid n>$\ are eigenvectors of $\hat{H}$,
\begin{eqnarray}
&&\hat{H}\mid n>=E_n\mid n> \label{w2}\\
&&E_n=H\left(\sum_{k=1}^Nm_kn_k+\sum_{k=1}^Nm_k\sigma_k\right) \nonumber
\end{eqnarray}
We see that the energy spectrum is degenerate: all eigenvectors $\mid n>$\ with $\sum_{k=1}^Nm_kn_k$\
 fixed give the same energy.
To classify the degeneracy according to the representations of some symmetry algebra we make the following trick \cite{b5}.
Let $m$\ be the least common multiple of all $m_k,\;k=1,\;\dots ,\;N$, and let $l_k=\frac{m}{m_k}$. Moreover, let us put
\begin{eqnarray}
n_k=q_kl_k+r_k,\; 0\leq r_k\leq l_k-1,\; k=1,\;\dots ,\;N\label{w3}
\end{eqnarray}

Then the expression (\ref{w2}) for energy eigenvalues takes the form
\begin{eqnarray}
E_n=H\left(m\sum_{k=1}^Nq_k+\sum_{k=1}^Nm_k(r_k+\sigma_k)\right)\label{w4}
\end{eqnarray}
Let $r\equiv (r_1,\;\dots ,\;r_N)$\ be fixed; define $X_r$\ to be the subspace spanned by all vectors $\mid n>$\ such that
$n_k=q_kl_k+r_k$; within $X_r$\ one can make an identification: $\mid n >\equiv \mid q>,\;q\equiv (q_1,\;\dots ,\;q_N)$.
The whole space of states is then the orthogonal sum of $l_1\cdot l_2\cdot \dots \cdot l_N$\ subspaces $X_r$,
\begin{eqnarray}
X =\oplus_rX_r\label{w5}
\end{eqnarray}

When restricted to $X_r$\ energy spectrum of $\hat{H}$\ takes particulary simple form described by eq. (\ref{w4}). The
relevant symmetry algebra can be constructed as follows. First, we define the creation and annihilation operators 
$a^+_i,\;a_i,\;i=1,\;\dots ,\;N$, as follows
\begin{eqnarray}
&&a^+_i\mid(n_1,\;\dots ,\;n_i,\;\dots ,\;n_N)>=\sqrt{n_i+1}\mid(n_1,\;\dots ,\;n_i+1,\;\dots ,\;n_N)> \nonumber \\
&&a_i\mid(n_1,\;\dots ,\;n_i,\;\dots ,\;n_N)>=\sqrt{n_i}\mid(n_1,\;\dots ,\;n_i-1,\;\dots ,\;n_N)> \label{w6}
\end{eqnarray}

On the other hand, in each $X_r$\ one can define another set of creation and annihilation operators
\begin{eqnarray}
&&b^+_{ri}\mid(q_1,\;\dots ,\;q_i,\;\dots ,\;q_N)>=\sqrt{q_i+1}\mid(q_1,\;\dots ,\;q_i+1,\;\dots ,\;q_N)> \nonumber \\
&&b_{ri}\mid(q_1,\;\dots ,\;q_i,\;\dots ,\;q_N)>=\sqrt{q_i}\mid(q_1,\;\dots ,\;q_i-1,\;\dots ,\;q_N)> \label{w7}
\end{eqnarray}
Obviously, the operators $b^+_{ri},\;b_{ri}$\ are expressible in terms of basic ones $a^+_i,\;a_i$; the relevant formulae read:
\begin{eqnarray}
&&b_{ri}=\left(\frac{N_i+l_i-r_i}{l_i}\right)^{1\over 2}\Pi_{s=1}^{l_i}(N_i+s)^{-\frac{1}{2}}\cdot a_i^{l_i}
\nonumber \\
&&b_{ri}^+=\left(\frac{N_i-r_i}{l_i}\right)^{1\over 2}\Pi_{s=1}^{l_i}(N_i-s+1)^{-\frac{1}{2}}\cdot (a_i^+)^{l_i}\label{w8}
\end{eqnarray}
where $N_i\equiv a_i^+a_i$\ are the standard particle-number operators. \\
Now, the operators $b_{ri},\;b_{ri}^+$\ can be used in order to build symmetry algebra in each $X_r$.
To this end let $\lambda_{\alpha},\;\alpha =1,\;\dots ,\;N^2$, form the basis of $u(N)$\ algebra.  Then the operators 
\begin{eqnarray}
\hat{\Lambda}_{r\alpha}\equiv \sum_{i,\;j=1}^Nb^+_{ri}(\lambda_{\alpha})_{ij}b_{rj}\label{w9}
\end{eqnarray}
have the following properties: (a) they are hermitean, (b) commute with $\hat{H}$\ in $X_r$, (c) obey $u(N)$\ commutation
rules in $X_r$.

Having constructed $u(N)$\ symmetry algebra in each $X_r$\ one can define the symmetry operators in the whole space of states $X$.
Let $Q_r$\ be the orthogonal projector on $X_r$. Then the operators
\begin{eqnarray}
\hat{\Lambda}_{\alpha}=\sum_rQ_r\hat{\Lambda}_{r\alpha}Q_r\equiv \sum_r\hat{\Lambda}_{r\alpha}Q_r\label{10}
\end{eqnarray}
span the $u(N)$\ symmetry algebra in the total space of states $X$.

It remains to construct the projectors $Q_r$. It is easy to check that the following operators do the job 
\begin{eqnarray}
&&Q_r\equiv \Pi_{i=1}^NP_i \label{w11}\\
&&P_i=\frac{1}{l_i}\sum_{s=0}^{l_i-1}e^{\frac{2i\pi s}{l_i}(N_i-r_i)}\nonumber
\end{eqnarray}

The creation and anihilation operators used above were constructed in a quite general abstract way.
However, an interesting question is whether they can be expressed in terms of basic dynamical variables.
In most cases the answer to this question is positive although the resulting expressions are, in general,
very complicated. Indeed, in almost all interesting cases (like, for example, interacting particles, with or without spin,
described by the natural hamiltonian ) the space of states carries an irreducible representation of the algebra
of observables. The only possible exceptions we can see are the existence of superselection rules or 
the cooexistence of discrete and continuum spectrum of the hamiltonian. 
In the latter case we must restrict ourselves to the discrete part of spectrum.
The basic dynamical variables do have, in general, nonvanishing matrix elements between eigenstates coresponding 
to discrete and continuous spectra. This can be cured, for example, by inserting the projector on discrete 
subspace but this makes the whole procedure even more complicated.

As we have mentioned above we are convinced that our assumptions (i)- (iii) are flexible enough to cover known
cases of superintegrable systems. Below we give a number of examples.

We start with the most obvious one - the harmonic oscillator with rational frequency ratios. 
The relevant hamiltonian reads 
\begin{eqnarray}
H=\sum_{k=1}^N ( \frac {p_k^2}{2m_k} + \frac {m_k^2l_k^2\omega ^2}{2}x_k^2 )=\sum_{k=1}^N H_k ,\label{w12}
\end{eqnarray} 

Defining 
\begin{eqnarray}
\hat{I_k}=\frac {1}{\hbar\omega l_k}H_k  \label{w13}
\end{eqnarray} 
we easily see that (i) - (iii) hold with all $\sigma _k=\frac {1}{2}$.

We can generalize this by adding the inverse square term \cite{b6}
\begin{eqnarray}
H=\sum_{k=1}^N ( \frac {p_k^2}{2m_k} + \frac {m_k^2l_k^2\omega ^2}{2}x_k^2 + \frac{\alpha _k^2}{x_k^2} )
=\sum_{k=1}^N H_k ;\label{w14}
\end{eqnarray} 

then
\begin{eqnarray}
\hat{I_k}=\frac {1}{2\hbar\omega l_k}H_k  \label{w15}
\end{eqnarray}
and $\sigma _k=\frac{3}{4}$\ for all k do the job.

The next example is the twodimensional Kepler problem 
\begin{eqnarray}
H=\frac {\vec p^2 }{2\mu } - \frac{\alpha }{r}  \label{w16}
\end{eqnarray}
The spectrum is descrete for negative energies and reads
\begin{eqnarray}
E_n=\frac {-\alpha \mu ^2}{2\hbar^2(n+\mid m\mid +1)^2},  \label{w17}
\end{eqnarray}
m being the value of angular momentum. Therefore, we restrict ourselves to the subspace spanned by negative-energy 
eigenvectors of (16). Moreover, due to the appearance of $\mid m\mid $\ in eq.(17) we split this subspace
futher accordingly to whether m is nonnegative or negative.

The dynamical symmetry behind is obtained by defining Runge-Lenz vector \cite{b7} 
\begin{eqnarray}
A_i=\frac {1}{2\mu }\varepsilon _{ik}(p_kL + Lp_k) - \frac{\alpha }{r}x_i . \label{w18}
\end{eqnarray}
One finds the following algebra 
\begin{eqnarray}
&& [L,\;A_i] = i\hbar\varepsilon _{ik}A_k \\  \label{w19}
&& [A_i,\;A_j] = - \frac{2i\hbar}{\mu }\varepsilon _{ij}HL \nonumber 
\end{eqnarray} 
This is a quadratic algebra. However, in the subspace of constant negative energy it becomes $sO(3)$\ algebra.

In order to see that (i) - (iii) are fulfiled consider the subspace corresponding to $ m\geq 0 $\ (the case $m < 0$\
can be treated along the same lines). Define
\begin{eqnarray}
&&\hat{I_1}=\frac {1}{\hbar}L \\ \label{w20}
&&\hat{I_2}=\frac{1}{\hbar}(\mu \sqrt{\frac{-\alpha }{2H }} - L ) \nonumber
\end{eqnarray}
These operators are well-defined, selfadjoint and commuting; their spectra obey (i) with $\sigma _1=0,\;\sigma _2=\frac{1}{2}$.
Moreover, 
\begin{eqnarray}
H=\frac {-\alpha \mu ^2}{2\hbar^2(\hat{I_1} + \hat{I_2})^2}  \label{w21}
\end{eqnarray}
so that $m_1=m_2=1$\ in eq.(1).

The threedimensional Kepler problem can be deal with in the same way. The orbital angular momentum $ \vec L $, together
with the the Runge - Lenz vector \cite{b7}
 
\begin{eqnarray}
\vec A = \frac{1}{2\mu }(\vec p\times \vec L-\vec L\times \vec P) - \frac{\alpha }{r}\vec r \label{w22}
\end{eqnarray}

form the quadratic algebra
\begin{eqnarray}
&&[L_i,\;L_j] = i\hbar\varepsilon _{ijk}L_k \\ \label{w23}
&&[L_i,\;A_j] = i\hbar\varepsilon _{ijk}A_k  \nonumber \\
&&[A_i,\;A_j] = -\frac{2ihbar}{\mu }\varepsilon {ijk}L_kH \nonumber
\end{eqnarray}
Again, let us consider the negative energy subspace. Define

 \begin{eqnarray}
&&\hat{I_1} = \sqrt{\frac{\vec L^2}{\hbar^2} + \frac{1}{4}} + \frac{L_3}{\hbar} \\ \label{w24}
&&\hat{I_2} =  \sqrt{\frac{\vec L^2}{\hbar^2} + \frac{1}{4}} - \frac{L_3}{\hbar}   \nonumber \\
&&\hat{I_3} = \frac{\mu }{\hbar }\sqrt{-\frac{\alpha }{2H}} - \frac{1}{2}(\hat{I_1}+\hat{I_2}) \nonumber
\end{eqnarray}
Then (i) - (iii) are obeyed with $\sigma _1 = \sigma _2 = \sigma _3 = \frac{1}{2} $\ and
  
 \begin{eqnarray}
 H = - \frac{2\alpha \mu ^2}{\hbar^2(\hat{I_1} + \hat{I_2} + 2\hat{I_3})^2} \label{w25}
\end{eqnarray}

Both harmonic oscillator and Kepler problem admit generalization consisting in replacing the Euclidean configuration
space by the sphere \cite{b8} which preserves superintegrability. It is easy to check using the results of \cite{b8}
that the relevant operators $\hat{I_k}$\ can be constructed in a similar way as in the "plane" case. Therefore, both
models fit into our scheme. 

Let us note in passing that our method applies to both models in arbitrary number of dimensions. 
This is because the dynamical
symmetries are here $SU(n)$\ or $ SO(n+1)$\ and one can use the Gelfand-Tseytlin method to construct the operators 
$\hat{I_k}$.

As a final example let us take the rational Calogero model \cite{b9}. It is known to be superintegrable also on 
the quantum level \cite{b10}. Using the results of \cite{b11} it is not difficult to check that our assumptions 
(i) - (iii) can be satisfied by an appropriate choice of $ \hat{I_k}$\ operators.

\end{document}